\documentclass[11pt,superscriptaddress,aps,prd,preprint]{revtex4}

\everymath{\displaystyle}
\usepackage{graphicx}
\usepackage{amsmath,amssymb,amsfonts,amstext,latexsym,stmaryrd,amscd,txfonts,pxfonts}

\begin{document}

\title{On Quantization of a Slow Rotating Kerr Black Hole }

\author{S. C. Ulhoa}
\email{sc.ulhoa@gmail.com}
\affiliation{Instituto de F\'isica, Universidade de Bras\'ilia, 70910-900, Bras\'ilia, DF, Brazil}

\author{E. P. Spaniol}
\email{spaniol.ep@gmail.com} \affiliation{UDF Centro
Universit\'ario and Centro Universit\'ario de Bras\'ilia UniCEUB, Bras\'ilia, DF, Brazil.}

\author{R. G. G. Amorim}
\email{ronniamorim@gmail.com}
\affiliation{Faculdade Gama, Universidade de Bras\'ilia, 70910-900, Bras\'ilia, DF, Brazil}

\begin{abstract}
In this article we calculate the total angular momentum for Kerr space-time for slow rotations. In order to analyze the role of such quantity we apply Weyl quantization method to obtain a quantum equation for the z-component of the angular momentum and for the squared angular momentum as well. We present an approximated solution by means the Adomian method. In such a method we find out a discrete angular momentum.
\end{abstract}
\maketitle

\section{Introduction} \label{sec.1}

In the old quantum theory the discrete nature of physical systems played an important role. Such a feature evolved giving rise to the so called quantum mechanics. In this process the passage of a classical theory to its quantum counterpart was developed by quantization techniques. Perhaps one of the first of such rules was the Bohr-Sommerfeld expression~\cite{bohr, bohr2, bohr3, sommerfeld, fedak} $$\oint p_\mu dx^\mu=n\hbar\,.$$ In fact such a quantization rule was applied to the Hydrogen spectral lines to obtain an atomic model. Hence the quantization of angular momentum in Bohr's model was fundamental to the establishment of the whole quantum mechanics. In this sense quantum theory was so successful explaining experimental data that nowadays it is considered one of the fundamental branches of theoretical physics. On the other hand at the macroscopic level where gravitational field is relevant when compared to other forces general relativity is equally successful. Both of such branches, general relativity and quantum mechanics, are considered pillars of modern physics. Yet they are mutually excluding each other. Thus the search for a quantum theory of gravitation is part of such an effort to unify all fields of physics. Although general relativity is the mainstream gravitational theory, it fails when one tries to apply quantization methods. This happens mainly because the definition of a gravitational energy in general relativity is problematic, in fact it is not possible to establish a quantity invariant under coordinates transformation and dependent on the reference frame in terms of the metric tensor.

In search for alternative theories of gravitation that recover all that success of general relativity, we found teleparallel gravity. It is a theory dynamically equivalent to general relativity which has been introduced by Einstein as an attempt to construct an unified field theory~\cite{einstein}. In the framework of teleparallel gravity there is a well defined expression for a gravitational energy-momentum tensor which also defines a gravitational angular momentum~\cite{maluf,maluf1}. However such an expression is not defined in phase space which demands a great effort to apply the canonical quantization rules. This forces one to use alternative methods such as Dirac method or Weyl's quantization~\cite{weyl,dirac}. Particularly the Weyl's method was used to obtain a discrete spectrum of mass for Schwarzschild space-time using teleparallel energy~\cite{advances}. However a quantization of gravitational angular momentum is still lacking. In this article we want to explore such a calculation for a slow rotating Kerr space-time.

This article is divided as follows. In section \ref{sec.2} the teleparallel gravity is described, the gravitational energy-momentum is introduced as well as the gravitational angular momentum. In section \ref{sec.3} we apply a quantization technique to gravitational angular momentum to obtain an eigenvalue equation for the respective operator and its square. We then give an approximated solution using the Adomian method. Finally in the last section we present our last comments. In this article we use natural unities unless otherwise stated. 

\section{Teleparallel Gravity} \label{sec.2}

Teleparallel gravity is an alternative theory of gravitation dynamically equivalent to general relativity. It's constructed out of tetrad field rather than metric tensor. The tetrad field relates two symmetries in space-time, Lorentz transformations and passive coordinate transformations. In order to tell them apart we use latin indices $a=(0),(i)$ to designate SO(3,1) symmetry and greek indices to diffeomorphism, $\mu=0,i$. Thus
\begin{eqnarray}
g^{\mu\nu}&=&e^{a\mu}e_{a}\,^{\nu}\,; \nonumber\\
\eta^{ab}&=&e^{a\mu}e^{b}\,_{\mu}\,, \label{1}
\end{eqnarray}
where $\eta^{ab}=diag(-+++)$ is the metric tensor of Minkowski
space-time. This means that for every metric tensor there are infinity tetrads, each of them is adapted to a specific reference frame. Teleparallel gravity is not only formulated in terms of tetrad fields but it is also defined in a Weitzenb\"ock geometry. Let's see how the equivalence to general relativity is obtained.

A weitzenb\"ockian manifold is endowed with the Cartan
connection~\cite{cartan},
$\Gamma_{\mu\lambda\nu}=e^{a}\,_{\mu}\partial_{\lambda}e_{a\nu}$,
which has a vanishing curvature tensor. On the other hand the torsion associated to such a connection is

\begin{equation}
T^{a}\,_{\lambda\nu}=\partial_{\lambda} e^{a}\,_{\nu}-\partial_{\nu}
e^{a}\,_{\lambda}\,. \label{4}
\end{equation}

The Christoffel symbols ${}^0\Gamma_{\mu \lambda\nu}$ are torsion free and exist in a riemannian geometry, thus the curvature tensor plays all dynamical roles for metric theories of gravitation such as general relativity. It is interesting to note that Christoffel symbols are related to Cartan connection by the following mathematical identity

\begin{equation}
\Gamma_{\mu \lambda\nu}= {}^0\Gamma_{\mu \lambda\nu}+ K_{\mu
\lambda\nu}\,, \label{2}
\end{equation}
where $K_{\mu \lambda\nu}$ is given by

\begin{eqnarray}
K_{\mu\lambda\nu}&=&\frac{1}{2}(T_{\lambda\mu\nu}+T_{\nu\lambda\mu}+T_{\mu\lambda\nu})\,,\label{3}
\end{eqnarray}
with $T_{\mu \lambda\nu}=e_{a\mu}T^{a}\,_{\lambda\nu}$, the quantity $K_{\mu\lambda\nu}$ is the contortion tensor.

The curvature tensor obtained from $\Gamma_{\mu \lambda\nu}$ is
identically zero which, using (\ref{2}), leads to

\begin{equation}
eR(e)\equiv -e({1\over 4}T^{abc}T_{abc}+{1\over
2}T^{abc}T_{bac}-T^aT_a)+2\partial_\mu(eT^\mu)\,,\label{eq5}
\end{equation}
where $R(e)$ is the scalar curvature of a Riemannian manifold and $T^\mu=T^b\,_b\,^\mu$. Since the
divergence term in eq. (\ref{eq5}) does not contribute with the
field equations, hence the Teleparallel Lagrangian density equivalent to Hilbert-Einstein Lagrangian density is

\begin{eqnarray}
\mathfrak{L}(e_{a\mu})&=& -\kappa\,e\,({1\over 4}T^{abc}T_{abc}+
{1\over 2} T^{abc}T_{bac} -T^aT_a) -\mathfrak{L}_M\nonumber \\
&\equiv&-\kappa\,e \Sigma^{abc}T_{abc} -\mathfrak{L}_M\;, \label{6}
\end{eqnarray}
where $\kappa=1/(16 \pi)$, $\mathfrak{L}_M$ is the Lagrangian
density of matter fields and $\Sigma^{abc}$ is given by

\begin{equation}
\Sigma^{abc}={1\over 4} (T^{abc}+T^{bac}-T^{cab}) +{1\over 2}(
\eta^{ac}T^b-\eta^{ab}T^c)\;, \label{7}
\end{equation}
with $T^a=e^a\,_\mu T^\mu$. The field equations obtained from such a Lagrangian read

\begin{equation}
\partial_\nu\left(e\Sigma^{a\lambda\nu}\right)={1\over {4\kappa}}
e\, e^a\,_\mu( t^{\lambda \mu} + T^{\lambda \mu})\;, \label{10}
\end{equation}
where $T^{\lambda \mu} $ is the energy-momentum of matter fields while $t^{\lambda\mu}$ which is defined by

\begin{equation}
t^{\lambda \mu}=\kappa\left[4\,\Sigma^{bc\lambda}T_{bc}\,^\mu- g^{\lambda
\mu}\, \Sigma^{abc}T_{abc}\right]\,, \label{11}
\end{equation}
represents the gravitational energy-momentum~\cite{maluf2}. It should be noted that $\Sigma^{a\lambda\nu}$ is skew-symmetric in the last two
indices, that leads to

\begin{equation}
\partial_\lambda\partial_\nu\left(e\Sigma^{a\lambda\nu}\right)\equiv0\,.\label{12}
\end{equation}
Therefore the total energy-momentum contained in a
three-dimensional volume $V$ of space is

\begin{equation}
P^a = \int_V d^3x \,e\,e^a\,_\mu(t^{0\mu}+ T^{0\mu})\,, \label{14}
\end{equation}
or using the field equations we have
\begin{equation}
P^a =4k\, \int_V d^3x \,\partial_\nu\left(e\,\Sigma^{a0\nu}\right)\,. \label{14.1}
\end{equation}
It worths to mention that the above expression is independent of coordinate transformations which is expected from a reliable definition of energy and momentum. On the other hand it is a vector under Lorentz transformations which is a feature of special relativity and there is no good reason to abandon such an attribute in gravitational theory.

We stress out the fact that the tetrad field is the dynamical variable of teleparallel gravity, then the usual definition of angular momentum in terms of the energy-momentum vector yields

\begin{equation}
L^{ab}= 4k\, \int_V d^3x \,e\left(\Sigma^{a0b}-\Sigma^{b0a}\right)\,,\label{15}
\end{equation}
this expression is the total angular momentum. Both $L^{ab}$ and $P^{a}$ obey a Poincar\'e algebra~\cite{ulhoa0} which is a very good indication of the consistency of the definition (\ref{15}).

\section{Angular Momentum Quantization} \label{sec.3}
\indent

The most general form of the line element that exhibits axial symmetry is given by

\begin{equation}
ds^2=g_{00}dt^2+2g_{03}d\phi
dt+g_{11}dr^2+g_{22}d\theta^2+g_{33}d\phi^2 \,. \label{5.1}
\end{equation}

That yields the following contravariant  metric tensor

\begin{equation}
g^{\mu\nu}=\left(%
\begin{array}{cccc}
  -\frac{g_{33}}{\delta} & 0 & 0 & \frac{g_{03}}{\delta} \\
  0 & \frac{1}{g_{11}} & 0 & 0 \\
  0 & 0 & \frac{1}{g_{22}} & 0 \\
  \frac{g_{03}}{\delta} & 0 & 0 & -\frac{g_{00}}{\delta} \\
\end{array}%
\right)  \label{29.2}
\end{equation}
where $\delta=g_{03}g_{03}-g_{00}g_{33}$. it should be noted that the components of the metric tensor are function of  $r$ e $\theta$.

In order to calculate the angular momentum we have to choose a referencial frame. Thus a tetrad field adapted to a stationary reference frame is given by
\begin{equation}
e_{a\mu}=\left(%
\begin{array}{cccc}
  -A & 0 & 0 & -B \\
  0 & \sqrt{g_{11}}\sin\theta\cos\phi & \sqrt{g_{22}}\cos\theta\cos\phi & -C\sin\theta\sin\phi \\
  0 & \sqrt{g_{11}}\sin\theta\sin\phi & \sqrt{g_{22}}\cos\theta\sin\phi & C\sin\theta\cos\phi \\
  0 & \sqrt{g_{11}}\cos\theta & -\sqrt{g_{22}}\sin\theta & 0 \\
\end{array}%
\right) \,, \label{5.2}
\end{equation}
with
\begin{eqnarray}
A &=& \sqrt{(-g_{00})} \,,
\nonumber \\
AB &=& -g_{03} \,, \nonumber \\
C\sin\theta &=& \frac{\delta^{1/2}}{\sqrt{(-g_{00})}} \label{29.31}
\end{eqnarray}
Then the non-vanishing components of the torsion tensor are
\begin{eqnarray}\nonumber
T_{013}&=& -A\partial_1B \,, \nonumber \\
T_{023}&=& -A\partial_2B \,, \nonumber \\
T_{001}&=& \frac{1}{2}\partial_1(A^2) \,, \nonumber \\
T_{002}&=& \frac{1}{2}\partial_2(A^2) \,, \nonumber \\
T_{112}&=& -\frac{1}{2}\partial_2(g_{11}) \,, \nonumber \\
T_{212}&=& \frac{1}{2}\partial_1(g_{22})-\sqrt{g_{11}g_{22}} \,, \nonumber \\
T_{313}&=& \frac{1}{2}\partial_1(g_{33})-\sqrt{g_{11}}C\sin^2\theta \,, \nonumber \\
T_{323}&=&\frac{1}{2}\partial_2(g_{33})-\sqrt{g_{22}}C\sin\theta\cos\theta\,.\label{torcaoaxial}
\end{eqnarray}

From expression (\ref{15}) it is possible to define the angular momentum density $M^{a b} =4ke\left(\Sigma^{a0b}-\Sigma^{b0a}\right)$, then after some algebraic manipulations it yields
\begin{equation}
M^{a b} = 2k \partial_i [e(e^{ai}e^{b0} - e^{bi}e^{a0})] \,,\label{Mab}
\end{equation}
which is an incredible simple expression in terms of the tetrad field adapted to a stationary reference frame.

The most historical representative of axial symmetry is the Kerr solution. Perhaps it is a natural step to investigate the quantization of angular momentum in such a system. In order to simplify our analysis let us take a slow rotating Kerr space-time which is given by the following line element 
\begin{equation}
ds^2 = - \Big( 1 - \frac{2m}{r} \Big)dt^2 + \Big( 1 - \frac{2m}{r}
\Big)^{-1}dr^2 + r^2d\theta^2 + r^2\sin^2 \theta d\phi^2 -
\frac{2ma}{r}\sin^2 \theta dtd\phi \,,\label{slowlyKerr}
\end{equation}
then the component of angular momentum in z-direction, $M_z = M_{(1)(2)}$, reads
\begin{equation}
M_z = \frac{ma \sin \theta}{r^2}\left( 1 - \frac{2m}{r} \right)^{-3/2}
\left[ - m \sin^2 \theta + r(3\cos^2 \theta -1)\sqrt{1 -
\frac{2m}{r}} \right]\ \,.\label{Mz}
\end{equation}

Similarly the modulus of the angular momentum, $M^2 = M^{ab}M_{ab}$, is given by

\begin{equation}
M^2 =\frac{m^2a^2 \sin^2 \theta}{r(8m^3 - r^3 + 6mr^2
-12m^2r)}\Big[r(2m-r)(4 - 3\sin^2 \theta)-m(2r+m)
\Big(1 - \frac{2m}{r} \Big)^{1/2}\sin^2 \theta \Big]\,.\label{M2}
\end{equation}

Now we have to apply some quantization procedure to those expressions of angular momentum. Thus the Weyl quantization is a mapping that lead classical coordinates, $z_n$, into operators $\widehat{z}_n$. Such a map is explicitly given by
\begin{equation}\label{w1}
\mathcal{W}[f(z_1, z_2,..., z_n)]:= \frac{1}{(2\pi)^n}\int d^n k d^n z f(z_1, z_2,..., z_n)\exp{\left(i\sum_{l=1}^{n} k_l(\widehat{z}_l - z_l)\right)}\,,
\end{equation}
usually the operators $\widehat{z}_n $ obey a non-commutative relation as $$[\widehat{z}_i, \widehat{z}_j]=\beta_{ij}\,.$$
It should be noted that both $M_z$ and $M^2$ depend on the coordinates $r$ and $\sin\theta$, hence we introduce the representation $\mathcal{W}[r]=\widehat{r}=\beta \frac{\partial}{\partial x}$ and
$\mathcal{W}[\sin \theta]=\widehat{x}=x$, with $\beta_{12}=\beta$. Therefore applying the Weyl prescription to $\mathcal{W}[M_z]= \widehat{M}_z $ and requiring

$$
\widehat{M}_z \psi = \lambda \psi\,,
$$
then 
\begin{equation}
j\frac{d^2\psi}{dx^2} + (3x^3 - 2x -3\mu j)\frac{d\psi}{dx} +
(\frac{9}{2}x^2 - 2\mu x^3 + 2\mu x - 1)\psi = 0\,, \label{qe1}
\end{equation}
where $\mu = \frac{m}{\beta}$ and $j = \frac{\lambda}{\mu a}$.

Analogously the Weyl procedure applied to $M^2$ together with an equation of eigenfunction and eigenvalue yields
\begin{eqnarray}\nonumber
j^2\frac{\partial^4\psi}{\partial x^4} &-&6\mu
j^2\frac{\partial^3\psi}{\partial x^3} + (3x^4 - 4x^2 + 12\mu^2
j^2)\frac{\partial^2\psi}{\partial x^2} + (8\mu x^2 -
8j^2 \mu^3
-8\mu x^4 - 8x +12 x^3)\frac{\partial\psi}{\partial x} \\
&+&(\mu^2x^4 - 16\mu x^3 + 8\mu x +18x -4)\psi = 0 \,.\label{qe2}
\end{eqnarray}
In order to present a solution for equations (\ref{qe1}) and (\ref{qe2}) we'll use the Adomian Decomposition Method.

\subsection{Adomian Decomposition Method (ADM)}

The Adomian Decomposition Method (ADM) was developed in 1961 to solve frontier physical problems \cite{adm1}. The method shows excellent results in the study of nonlinear ordinary differential, integro-differential and partial differential equations. It is based on the following steps: consider an equation 
\begin{equation}\label{adm1}
Hy(x)=f(x),
\end{equation} 
where $H$ stands for a general nonlinear ordinary differential operator, with linear and nonlinear terms. The linear part can be separated in two others, $L$ and $R$, where $L$ is easily inverted and $R$ is the remainder of the linear operator. In this sense, operator $H$ can be written as
$$H=L+R+N,$$
where $N$ is the nonlinear part. Then, equation (\ref{adm1}) becomes
\begin{equation}\label{adm2}
Ly(x)+Ry(x)+Ny(x)=f(x).
\end{equation}
Equation (\ref{adm2}) can be written as
\begin{equation}\label{adm3}
L^{-1}[Ly(x)]=-L^{-1}[Ry(x)]-L^{-1}[Ny(x)]+L^{-1}[f(x)],
\end{equation}
where $L^{-1}$ is the inverse operator of $L$. If $L$ is a second order operator, for example, we have $L^{-1}[Ly(x)]=y(x)-y(0)-y'(0)x$, and the solution of equation (\ref{adm3}) turns out to be
\begin{equation}\label{adm3}
y(x)=y(0)+y'(0)x-L^{-1}[Ry(x)]-L^{-1}[Ny(x)]+L^{-1}[f(x)].
\end{equation}
The nonlinear term $Ny(x)$ can be expanded as $\sum_{n=0}^{\infty}A_n$, where $A_n$ are the so called Adomian polynomials. The remaining part $y(x)$ will be decomposed into $y(x)=\sum_{n=0}^{\infty}y_n$, with $y_0=y(0)+y'(0)x+L^{-1}[f(x)]$. Consequently, we have
\begin{equation}\label{adm4}
y(x)=y_0-L^{-1}[R(\sum_{n=0}^{\infty}y_n)]-L^{-1}[\sum_{n=0}^{\infty}A_n].
\end{equation}
The Adomian polynomials can be calculated using the relation
\begin{equation}\label{adm5}
A_n=\frac{1}{n}\left[\frac{d}{d\lambda^n}[N(\sum_{n=0}^{\infty}\lambda^{i}y_{i})]\right]_{\lambda=0},
\end{equation}
$n=0,1,2,3,...$.
An extensive use of such a method can be found in reference \cite{adm1}.

If we apply this method to equation (\ref{qe1}), then we obtain

\begin{equation}\label{solqe1a}
\psi(x)=\psi(0)+\frac{d\psi(0)}{dx}x-\frac{1}{j}\left\{L^{-1}\left[(3x^3-2x-3\mu j)\frac{d\psi(x)}{dx}\right]-L^{-1}\left[(-2\mu x^3+\frac{9}{2}x^2+2\mu x-1)\psi(x)\right]\right\}.
\end{equation}
Hence using  $\psi(x)=\sum_{n=0}^{\infty}\psi_n$, we have
$$\psi_0(x)=\psi(0)+\frac{d\psi(0)}{dx}x.$$
With the boundary condition $\psi(1)=\psi(-1)$, we conclude that $\frac{d\psi(0)}{dx}=0$. Thus the first order approximation becomes
$$\psi_0(x)=\psi(0).$$ It should be pointed out that the boundary condition used has a clear meaning, it says that the information about angular momentum is the same at the poles of Kerr ergosphere. The superior orders of approximation in the solution can be found by an iterative procedure, it is given by
$$\psi_1(x)=-\frac{1}{j}\left\{L^{-1}\left[(3x^3-2x-3\mu j)\frac{d\psi_0(x)}{dx}\right]-L^{-1}\left[(-2\mu x^3+\frac{9}{2}x^2+2\mu x-1)\psi_0(x)\right]\right\},$$
$$\psi_2(x)=-\frac{1}{j}\left\{L^{-1}\left[(3x^3-2x-3\mu j)\frac{d\psi_1(x)}{dx}\right]-L^{-1}\left[(-2\mu x^3+\frac{9}{2}x^2+2\mu x-1)\psi_1(x)\right]\right\},$$
$$\psi_{n+1}(x)=-\frac{1}{j}\left\{L^{-1}\left[(3x^3-2x-3\mu j)\frac{d\psi_n(x)}{dx}\right]-L^{-1}\left[(-2\mu x^3+\frac{9}{2}x^2+2\mu x-1)\psi_n(x)\right]\right\}.$$
In this way, the solution of equation (\ref{qe1}) up to the second order approximation is 
\begin{eqnarray}\label{solqe1b}
\psi(x)&=&\psi(0)+\frac{\psi(0)}{j}\left[\frac{x^5}{10}-\frac{3x^4}{8}-\mu x^3+\frac{x^2}{2}\right]+\frac{\psi(0)}{j^2}\Biggl[\frac{\mu^2x^{10}}{450}-\frac{27\mu x^9}{720}+\left(\frac{99}{16}+\frac{9\mu^2}{5}\right)\frac{x^8}{56}+\frac{294 \mu x^7}{1680}\nonumber	\\
&-&\left(\frac{69}{8}+\frac{3\mu^2}{2}\right)\frac{x^6}{30}+\left(-\frac{21 \mu}{2}+2\mu j\right)\frac{x^5}{20}+\left(\frac{5}{2}-2\mu j-\frac{9j}{2}\right)\frac{x^4}{12}+\frac{\mu j x^3}{6}\Biggr].
\end{eqnarray}
With the condition $\psi(1)=\psi(-1)$, we estimate that the lower value for eigenvalue $j$ is $j=\sqrt{6}$. It should be noted that $j$ assumes discrete values depending on the approximation order taken in ADM.

Similarly the solution obtained when ADM is applied to (\ref{qe2}), again up to the second order approximation, reads
\begin{eqnarray}\label{solqe2}
\psi(x)&=&\psi(0)+\frac{d\psi(0)}{dx}x+\frac{d^2\psi(0)}{dx^2}x^2+\frac{1}{6}\frac{d^3\psi(0)}{dx^3}x^3\\\nonumber
&+&\left(6\frac{d^3\psi(0)}{dx^3}\mu-24\frac{d^3\psi(0)}{dx^3}\mu^2+8\frac{d\psi(0)}{dx}\mu^3+\frac{4}{j^2}\psi(0)\right)\frac{x^4}{4!}\\\nonumber
&+&\left(-12\frac{d^3\psi(0)}{dx^3}\mu^2+\frac{8}{j^2}\frac{d\psi(0)}{dx}-\frac{(18-8\mu)}{j^2}-\frac{4}{j^2}\frac{d\psi(0)}{dx}\right)\frac{x^5}{5!}\\\nonumber
&+&\left(\frac{12}{j^2}\frac{d^2\psi(0)}{dx^2}-\frac{8}{j^2}\frac{d\psi(0)}{dx}\mu-4\frac{d^3\psi(0)}{dx^3}\mu^3-\frac{(18-8\mu)}{j^2}\right)\frac{x^6}{6!}\\\nonumber
&+&\left(\frac{22}{3j^2}\frac{d^3\psi(0)}{dx^3}-\frac{6}{j^2}\frac{d^3\psi(0)}{dx^3}\mu-\frac{12}{j^2}\frac{d\psi(0)}{dx}+\frac{16}{j^2}\psi(0)\mu-\frac{(18+8\mu)}{j^2}\frac{d^2\psi(0)}{dx^2}\right)\frac{x^7}{7!}\\\nonumber
&+&\left(-\frac{30}{j^2}\frac{d^2\psi(0)}{dx^2}+\frac{5}{4j^2}\frac{d\psi(0)}{dx}\mu-\frac{1}{j^2}\psi(0)\mu^2-\frac{(18+8\mu)}{6j^2}\frac{d^3\psi(0)}{dx^3}\right)\frac{x^8}{8!}\\\nonumber
&+&\left(-\frac{3}{j^2}\frac{d^3\psi(0)}{dx^3}+\frac{32}{j^2}\frac{d^2\psi(0)}{dx^2}\mu-\frac{1}{j^2}\frac{d\psi(0)}{dx}\mu^2\right)\frac{x^9}{9!}\\\nonumber
&+&\left(\frac{20}{3j^2}\frac{d^3\psi(0)}{dx^3}\mu-\frac{1}{j^2}\frac{d^2\psi(0)}{dx^2}\mu^2\right)\frac{x^{10}}{10!}-\left(\frac{1}{6j^2}\frac{d^3\psi(0)}{dx^3}\mu^2\right)\frac{x^{11}}{11!}\,.\nonumber
\end{eqnarray}
This solution gives a discrete value of angular momentum as well as above. On the other hand we did not identify an explicit form of such discrete angular momentum.

\section{Conclusion} \label{sec.4}

In this article we have presented the quantization of angular momentum in a slow rotating Kerr space-time. The angular momentum was obtained in the realm of teleparallel gravity which separates features of gravitational field from matter fields. A slow rotating approximation was used to calculate the angular momentum and the Weyl quantization procedure was used to obtain a quantum equation. Such a prescription was applied to the z-direction of angular momentum and its squared. The equations were established by an eigenvalue-eigenfunction equation. We used the Adomian method to obtain an approximated solution and using boundary conditions we find out a discrete angular momentum eigenvalue. Such a discrete feature can help us to look for experimental evidences of a quantum theory of gravitation.


\begin{thebibliography}{99}
\bibitem{bohr}
N. Bohr, Philos. Mag. 26, 1 (1913).

\bibitem{bohr2}
N. Bohr, Philos. Mag. 26, 476 (1913).

\bibitem{bohr3}
N. Bohr, Philos. Mag. 26, 857 (1913).

\bibitem{sommerfeld}
A. Sommerfeld. ``Zur Quantentheorie der Spektrallinien''. Annalen der Physik. v. 356 (17): 1â94. 1916.

\bibitem{fedak}
W. A. Fedak and J. J. Prentis. The 1925 Born and Jordan paper ``On quantum mechanics''. American Journal of Physics - AMER J PHYS. 77, 2009.

\bibitem{einstein}
A.~Einstein.
\newblock {Auf die Riemann-Metrik und den Fern-Parallelismus gegr{\"{u}}ndete
  einheitliche Feldtheorie}.
\newblock {\em Math. Ann.}, 102(1):685--697, dec 1930.


\bibitem{maluf}
J. W. Maluf. Hamiltonian formulation of the teleparallel description of general relativity. Journal of Mathematical Physics, Nova York, EUA, v. 35, p. 335, 1994.

\bibitem{maluf1}
Jos{\'{e}}~W. Maluf.
\newblock {The teleparallel equivalent of general relativity}.
\newblock {\em Ann. Phys.}, 525(5):339--357, may 2013.

\bibitem{weyl}
Hermann Weyl.
\newblock {\em {The theory of groups and quantum mechanics}}.
\newblock Dover Publications, New York, 1931.

\bibitem{dirac}
P.~A.~M. Dirac.
\newblock {Generalized Hamiltonian Dynamics}.
\newblock {\em Proc. R. Soc. A Math. Phys. Eng. Sci.}, 246(1246):326--332, aug
  1958.

\bibitem{advances}
S.~C. Ulhoa and R.~G.~G. Amorim.
\newblock {On Teleparallel Quantum Gravity in Schwarzschild Space-Time}.
\newblock {\em Adv. High Energy Phys.}, 2014:1--6, may 2014.

\bibitem{cartan}
E. Cartan, On a generalization of the notion of Reimann curvature and spaces
with torsion, in NATO ASIB Proc. 58: Cosmology and Gravitation: Spin, Torsion,
Rotation, and Supergravity, eds. P. G. Bergmann and V. de Sabbata (1980), pp. 489–
491.

\bibitem{maluf2}
J. W. Maluf, J. F. Rocha Neto, T. M. L. Tor\'ibio, K. H. C. Branco. Energy and angular momentum of the gravitational field in the teleparallel geometry. Physical review. D, Particles, Fields, Gravitation, and Cosmology, Estados Unidos, v. 65, n.12, p. 124001, 2002.

\bibitem{ulhoa0}
J. W. Maluf, S. C. Ulhoa, F. F. Faria, J. F. Rocha Neto. The angular momentum of the gravitational field and the Poincar\'e group. Classical and Quantum Gravity, v. 23, p. 6245-6256, 2006.

\bibitem{adm1} 
G. Adomian, \emph{Solving Frontier Problems of Physics:}The Decompostion Method, Kluwer Academic, Boston, 1994.
	
\end{thebibliography}

\end{document}